\documentclass{aa}
\usepackage{lineno}
\usepackage{booktabs}

\usepackage{graphicx}
\usepackage[varg]{txfonts}
\usepackage{natbib}
\bibpunct{(}{)}{;}{a}{}{,}
\usepackage{array}
\usepackage{xspace}
\usepackage{lscape}
\usepackage{url}
\usepackage{arydshln}
\usepackage[hyperfootnotes=false, linktocpage=true, breaklinks=true, colorlinks=true, linkcolor=blue, citecolor=blue, urlcolor=blue]{hyperref}
\usepackage[all]{hypcap}
\usepackage{longtable}
\usepackage{afterpage}
\usepackage[dvipsnames]{xcolor}
\usepackage{xfrac}
\usepackage{underscore}
\usepackage{amsmath}

\mathchardef\mhyphen="2D
\begin{document}

\title{Delving into the depths of NGC 3783 with XRISM}
\subtitle{VI. Evidence for a failed wind based on \ion{Fe}{XXV} absorption line variability }

\author{
Chen Li \inst{1,2}
\and
Jelle S. Kaastra \inst{2,1}
\and
Liyi Gu \inst{2,1}
\and
Missagh Mehdipour \inst{3}
\and
Keigo Fukumura \inst{4}
\and
Matteo Guainazzi \inst{5}
\and
Misaki Mizumoto \inst{6}
\and
Ehud Behar \inst{7, 8}
\and
Shoji Ogawa \inst{9}
\and
Keqin Zhao \inst{1, 2}
}
\institute{
Leiden Observatory, Leiden University, PO Box 9513, 2300 RA Leiden, the Netherlands  \\ \email{cli@strw.leidenuniv.nl}
\and
SRON Netherlands Institute for Space Research, Niels Bohrweg 4, 2333 CA Leiden, the Netherlands
\and
Department of Astronomy, University of Michigan, 1085 South University Avenue, Ann Arbor, MI 48109, USA
\and
Department of Physics and Astronomy, James Madison University, Harrisonburg, VA 22807, USA
\and
ESA European Space Research and Technology Centre (ESTEC), Keplerlaan 1, 2201 AZ, Noordwĳk, the Netherlands
\and
Science Research Education Unit, University of Teacher Education Fukuoka, Munakata, Fukuoka 811-4192, Japan
\and
Department of Physics, Technion, Haifa 32000, Israel
\and
MIT Kavli Institute for Astrophysics and Space Research, Massachusetts Institute of Technology, Cambridge, MA 02139, USA
\and
Faculty of Science and Technology, Tokyo University of Science, 2641 Yamazaki, Noda, 2788510, Chiba, Japan
}
\abstract
{
NGC 3783 exhibits a prominent \ion{Fe}{XXV} absorption trough in the time-averaged XRISM/Resolve spectrum.
To investigate its short-term variability, we present a time-resolved analysis of the \ion{Fe}{XXV} absorption line during the continuous 10-day observation. 
We divided the resolved absorption profile into five fixed velocity intervals and measured their normalized equivalent widths in 19 time-resolved spectra.
Three velocity intervals exhibit significant equivalent-width (EW) variability. 
Two  velocity intervals respond differently to low- and high-flux states with less emission line contamination.
The time-dependent photoionisation modelling 
is unable to reproduce the most significant observed EW variation ($3.5 \, \sigma$), possibly owing to uncertainties in 
the assumed initial equilibrium state, which prevent a precise density measurement. 
Instead, using photoionisation calculations and assuming the high-density limit, we derive approximate constraints on the ionisation parameters of the two relevant components. The low-velocity component ($-320$ to $-680~\rm km\,s^{-1}$) shows a delayed inverse response, indicating a high ionisation parameter, while its relatively short lag timescale indicates a relatively high density and a distance
upper limit of \(0.05\) pc. In contrast, the high-velocity component ($-680$ to $-1040~\rm km \,s^{-1}$) manifests a positive response with a longer lag to the ionising continuum, implying a relatively low ionisation parameter.
This component appears to be less dense, with an inferred distance upper limit of 0.25 pc.
The low-velocity component is consistent with a failed wind, as its outflow velocity is below the local escape velocity, whereas the high-velocity component lies near the escape threshold. 
Together with the lower-ionisation failed wind detected in 2001, this supports a multiphase failed-wind scenario in the unobscured state of NGC~3783 over more than two decades.
}
\authorrunning{Li C., et al.}
\titlerunning{Variability of the \ion{Fe}{XXV} resonance absorption line in NGC 3783}
\maketitle

\nolinenumbers


\section{Introduction}
\label{sect_intro}
Active galactic nuclei (AGNs) commonly exhibit ionized outflows spanning a broad range of column densities, ionisation states, and velocities, from classical warm absorbers (WAs) to highly ionized Fe K absorbers detected in the X-ray band \citep{Tombesi2013, Yamada2024}. 
These outflows are closely linked to the accretion--ejection processes operating near the central supermassive black hole (SMBH), and are widely regarded as an important manifestation of AGN feedback \citep{Tombesi2010, Gaspari2017, Giustini2019, Laha2021}. 
In particular, highly ionized Fe K absorption lines provide a direct probe of the innermost circumnuclear gas, where the wind launching and acceleration processes are expected to occur \citep{Tombesi2013, Gofford2015}.

The Resolve microcalorimeter onboard XRISM delivers a spectral resolution of approximately 4.5 eV full width at half maximum (FWHM) at 6 keV \citep{Tashiro2025,Kelley2025}.
This high resolution enables precise measurements of highly ionized outflows in the Fe K band, such as \ion{Fe}{XXV} and \ion{Fe}{XXVI} lines, as demonstrated by early XRISM observations of several AGN, including PDS 456 (\citealp{PDS2025, Xu2025}), NGC 3783 (\citealp{Mehdipour2025}), IRAS 05189–2524 (\citealp{Noda2025}), NGC 4151 (\citealp{Xiang2025, Miller2026}), Mrk 279 (\citealp{Miller2025}), MCG$-$6$-$30$-$15 (\citealp{Brenneman2025}), PG 1211+143 (\citealp{Mizumoto2026}), NGC 3516 (\citealp{Juranova2026}), NGC 1365 (\citealp{Zaidouni2026}), Cen A (\citealp{Kayanoki2026}) and NGC 5548 (\citealp{Mehdipour2026}). 
These observations have revealed outflows with velocities ranging from a few hundred km s$^{-1}$ to a substantial fraction of the speed of light, spanning a wide range of radial distances and often exhibiting complex, clumpy structures.

Time-resolved analysis of XRISM/Resolve spectra provides a powerful means of diagnosing variability in Fe K absorbers on daily, and even sub-day, timescales.
It offers insight into the acceleration mechanism for a detected transient Ultra-fast Outflow (UFO) during a soft flare in NGC 3783 \citep{Gu2025}.
In NGC 4051, the Fe K absorber changes within less than a day, possibly because of two separate absorbers or one accelerating outflow \citep{Reeves2026}.
In NGC 3516, three highly ionized Fe K absorbers vary on timescales of tens of kiloseconds, likely due to changes in ionisation and/or clumps crossing the line of sight \citep{Juranova2026}.
Overall, Resolve time-resolved spectroscopy is a powerful tool for characterizing rapidly varying outflows under photoionisation-equilibrium models.

Time-dependent photoionisation modelling of AGN outflows, by diagnosing the response of the ionisation state of the absorbing outflows to changes in the ionizing continuum, uses the recombination timescale ($t_{\rm rec}$) of the outflow to constrain its electron number density ($n_{\rm e}$) by $n_{\rm e} \propto 1 / t_{\rm rec}$ \citep{Krolik&Kriss1995, Nicastro1999, Kaastra2012}.
The subsequent spectral-timing technique, which is based on the advanced time-dependent photoionisation models \citep{Rogantini2022, Sadaula2023}, has been used to measure or place a precise constraint on density/distance for low-ionisation WAs (\citealt{Li2023,Li2025,Sadaula2025}) with
ionisation parameter \citep{Tarter1969, Krolik&Kriss1995}
\begin{equation}\label{equ:xi defination}
      \xi = \frac{L_{\rm ion}}{n_{\rm H} r^{2}} \,  ,
\end{equation}
where $L_{\rm ion}$ is the 1$ - $1000 Ryd ionizing luminosity, $n_{\rm H}$ is the hydrogen density, $r$ is the distance of the gas to the ionizing source. 
One interesting resulting application is the identification of a failed wind candidate visible through the Fe Unresolved Transition Arrays (UTA, \citealt{Behar2001}) in NGC 3783 \citep{Li2025}.
Such constraints can help break the degeneracy among different launching mechanisms \citep{Fukumura2010, Fukumura2018, Fukumura2024, Takeuchi2013, Giustini2019, Giustini2021, Mizumoto2019, Mizumoto2021, Owen2025} and clarify the connection between UFOs and warm absorbers \citep{Tombesi2013,Laha2021}.

Here we present {a} study of the variability of the highly ionized Fe-K absorbers (\ion{Fe}{XXV} at 6.7 keV) from the first long observation of NGC 3783 led by XRISM in 2024.
NGC 3783 is one of the best-studied Seyfert 1 galaxies in high-resolution X-ray spectroscopy and has long served as a benchmark source 
exhibiting multiple-phase ionized AGN outflows with a broad range of velocities
\citep{Kaspi2001,Kaspi2002,Krongold2003,Netzer2003, Behar2003, Mao2019, Mehdipour2025, Zhao2026}. 
XRISM observed the source continuously for 10 days, and the corresponding time-averaged spectrum reveals a prominent \ion{Fe}{XXV} absorption feature around 6.7 keV composed of three different photoionized components \citep{Mehdipour2025}.
With CCD spectral resolution, the strong 6.7 keV absorption line is also clearly detected in the time-averaged XMM-Newton/PN spectra by \citet{Reeves2004}, while the depth of the feature appears to vary over a timescale of $10^5$s, being strongest when the continuum flux is higher.
In grating spectra, the \ion{Fe}{XXV} resonance line is hard to detect due to the {relatively} small effective area of Chandra/HETG \citep{Kaspi2002}.
With the unprecedented energy resolution and enhanced effective area in the Fe K band of Resolve, the time-averaged spectrum of NGC 3783 provides us with a high signal-to-noise ratio, especially important for studying the \ion{Fe}{XXV} band in the greatest detail.

The \ion{Fe}{XXV} absorption line of NGC 3783 is the most prominent and relatively broad absorption feature in the Fe-K band, enabling us to study the variability of the multiple kinematic components of this line.
Furthermore, using state-of-the-art photoionisation modelling in SPEX version 3.08.01 (\citealp{Kaastra1996, Kaastra2024}), the density and distance of the highly ionized \ion{Fe}{XXV} absorber are expected to be constrained more precisely, thereby advancing our understanding of its origin.
This paper is organized as follows. 
In Sect. \ref{sect_obs}, we describe the time-resolved analysis of the Resolve spectra of \ion{Fe}{XXV} and the derived variability for multiple kinematic velocity components. 
In Sect. \ref{sect3_model}, we model the detected variability with time-dependent photoionisation modelling (\textit{tpho}, \citealp{Rogantini2022}). 
We discuss and summarize our results in Sect. \ref{sect_disc} and Sect. \ref{sect_conclusion}, respectively.

\section{Observation data analysis and results}
\label{sect_obs}

NGC 3783 was observed by XRISM for $10$ days, from July $18-27$ in 2024.
The coordinated multiwavelength campaign also included XMM-Newton, Chandra, NuSTAR, Swift, NICER, and HST, see  \cite{XRISMn3783cal2025} for details.
In this work, we primarily use data from XRISM (Resolve $439\, \rm ks$, Xtend $372\, \rm ks$) and Swift/XRT (long-term monitoring with 30 data points) for our time-resolved spectra analysis and time-dependent photoionisation modelling. 
We adopt the same spectral dataset as in our previous NGC 3783 series (Papers I–V) and refer the reader to those papers for full details of the data reduction and analysis procedures of XRISM \citep{Mehdipour2025, XRISMn3783cal2025, Gu2025, Li2026, Zhao2026}.

\begin{figure*}[!tbp]
\centering
\hspace*{-0.00cm}\resizebox{1.0\hsize}{!}{\includegraphics[angle=0]{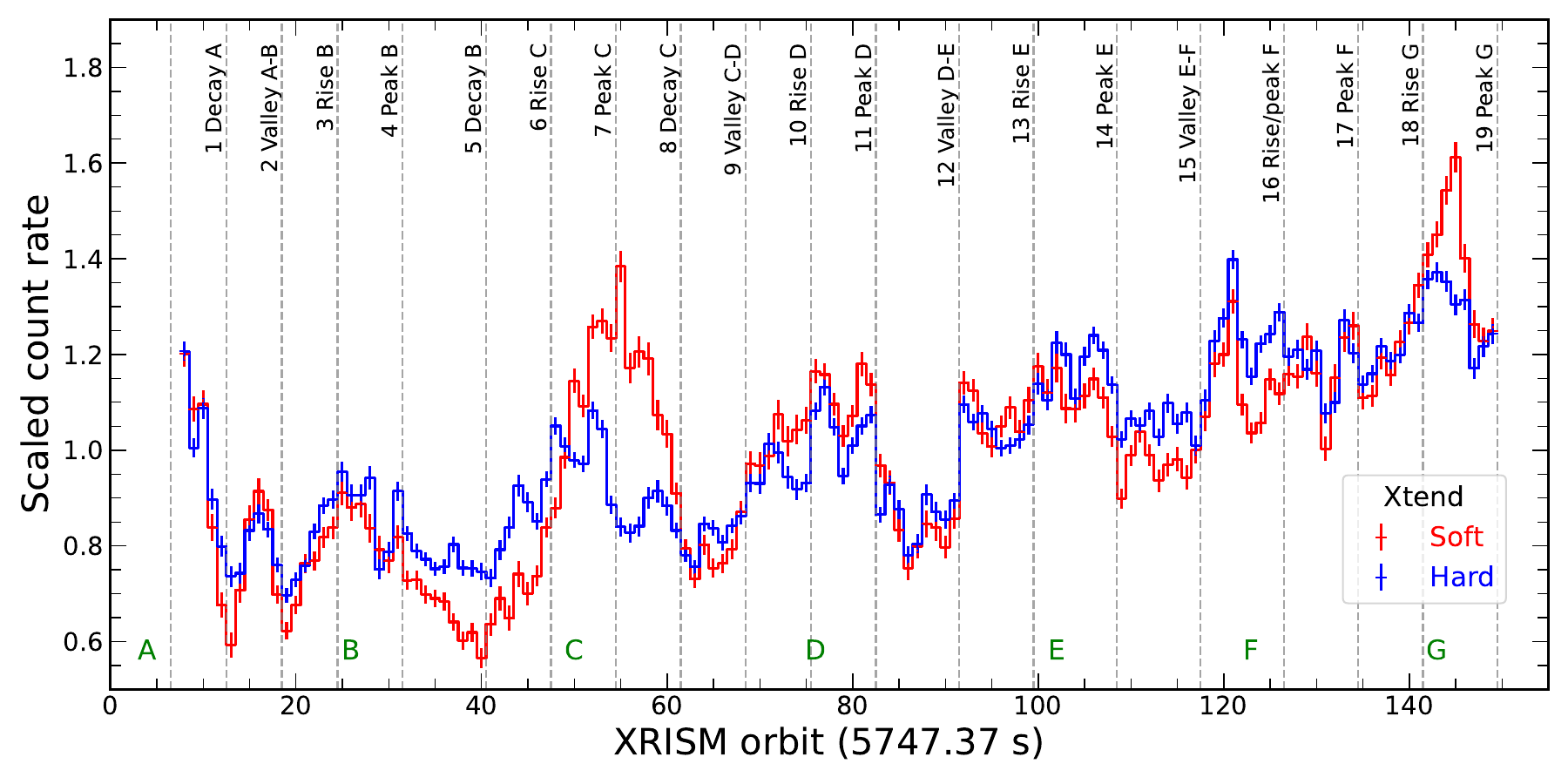}}\vspace{-0.0cm}
\caption{
XRISM/Xtend light curves of NGC~3783 in the soft
($0.3$--$0.7~\mathrm{keV}$; red) and hard
($3$--$6~\mathrm{keV}$; blue) bands. The vertical dashed lines delimit
the 19 time intervals used for the time-resolved spectral analysis, with
each interval labelled according to the local flux morphology. The letters
A--G mark the seven main variability episodes. One XRISM orbit corresponds
to $5747.37~\mathrm{s}$.
Orbit 0 corresponds to 2024 July 17 at 19:43:44 UTC, marking the start of the first Chandra observation in the 2024 XRISM campaign.
}
\label{fig_xtend_interval}
\end{figure*}

The main goal of this paper is to investigate the response of the \ion{Fe}{XXV} absorption line to the ionizing flux variations in NGC 3783.
Due to the closed gate valve of the XRISM/Resolve instrument, that instrument is insensitive to soft X-rays. Therefore, we present the XRISM/Xtend light curves in the 0.3–0.7 keV and 3.0–6.0 keV bands in the top panel of Fig. \ref{fig_xtend_interval} to illustrate the intrinsic source variability.
These light curves were used as a reference for defining the time intervals for time-resolved spectral analysis. 
The boundaries were placed at transitions between rise, peak, decay, and valley phases, while ensuring sufficient counts in the \ion{Fe}{XXV} line. 
All phases are resolved in the light curve and may sample different stages of the absorber response.
Seven major flux peaks can be distinguished (labelled A $-$ G in green); the strongest, here labelled "C", is the prominent soft-X-ray flare, which has been linked to a transient Ultra-Fast Outflow launching episode by \cite{Gu2025}.

The 19 time intervals shown in Fig.~\ref{fig_xtend_interval}, each lasting approximately half a day, {ensure a sufficient signal-to-noise ratio for investigating} variability in the \ion{Fe}{XXV} resonance absorption line.
Shorter intervals are not useful for this purpose, because the signal-to-noise ratio in the energy band of the absorption line becomes too poor. For each time interval, we {extracted} a Resolve spectrum, {yielding 19 time-resolved spectra for the analysis in Sect.~\ref{parameter_set-up}.}

\subsection{Local fit to the time-averaged Fe XXV velocity profile}
\label{sect:local_fit_to_averaged}

\begin{figure}[!tbp]
\centering
\hspace*{-0.00cm}\resizebox{1.05\hsize}{!}{\includegraphics[angle=0]{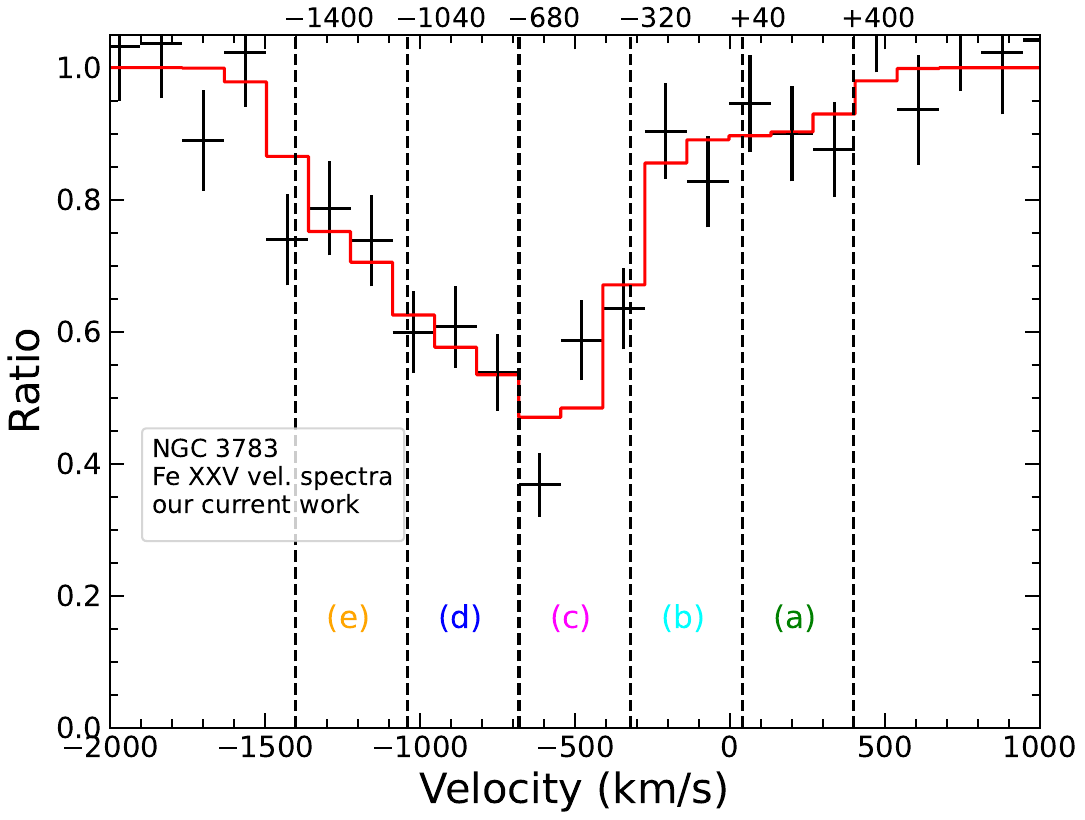}}\vspace{-0.0cm}
\caption{
The \ion{Fe}{XXV} velocity {profile} of NGC 3783 {derived} from the 439 ks time-averaged XRISM/Resolve spectrum {obtained} in 2024.
The data-to-model ratio (black) shows the residual \ion{Fe}{XXV} absorption relative to the baseline model of \cite{Mehdipour2025}, from which the \ion{Fe}{XXV} absorption components were removed.
The red line represents the local best-fit model of the local spectrum, convolved with the instrument, using the model with five piecewise-constant transmission components as described in the text.
}
\label{fig_5etaus}
\end{figure}

For the time-averaged \ion{Fe}{XXV} velocity profile shown in
Fig.~\ref{fig_5etaus}, we divided the central region and the red and blue
wings of the absorption profile into five adjacent energy bins, each
approximately $8~\mathrm{eV}$ wide and corresponding to a velocity interval
of $360~\mathrm{km\,s^{-1}}$.
The adopted width of 360 km/s for the five line components is a compromise between resolution and significance. Narrower components lead to strong correlations between the equivalent widths of neighbouring components, and, consequently, to poorly constrained line properties.

We used the spectral model of \citet{Mehdipour2025}  as our baseline, with one modification.
\cite{Mehdipour2025} modelled the time-averaged Resolve spectrum with a power-law continuum, emission-line components, and six photoionized \textit{xabs} transmission components in a global fit. 
Each \textit{xabs} \footnote{\url{https://spex-xray.github.io/spex-help/models/xabs.html}} component describes the time-averaged
properties of a corresponding warm-absorber component. The \ion{Fe}{XXV}
absorption trough studied here receives its principal contributions from
three of these components (see Sect.~\ref{tpho}). 
To sample the resolved asymmetric \ion{Fe}{XXV} absorption trough, we removed the \ion{Fe}{XXV} contributions of these three
\textit{xabs} components and modelled the absorption trough locally
in each velocity interval with the transmission, $T(E)$, constant and given by $T(E) = 1 - f_{\rm cov}$.

In the local fit to the time-averaged \ion{Fe}{XXV} spectrum, we kept the
parameters of all remaining model components adopted from
\citet{Mehdipour2025} unchanged. For each 
component, only the
covering factor $f_{\mathrm{cov}}$ was fitted. Under our adopted
opaque-bin configuration, $f_{\mathrm{cov}}$ is approximately equal to the
normalized equivalent width (EW, see Sect.~\ref{parameter_set-up}). The five 
components
therefore measure the absorption strength within the five fixed velocity
intervals. The normalized EWs derived from the time-averaged spectrum are
adopted as reference values for the corresponding intervals (see the second
row of Table~\ref{tab:19interval_stats}). The best-fitting local model is
shown by the red curve in Fig.~\ref{fig_5etaus}.

\begin{table*}
\centering
\caption{
Statistical properties of the \ion{Fe}{XXV}
equivalent-width variability in five velocity intervals.
}
\label{tab:19interval_stats}
\renewcommand{\arraystretch}{1.15}
\setlength{\tabcolsep}{6pt}
\begin{tabular}{l|c|c|c|c|c}
\hline
\hline
Velocity range & (a) & (b) & (c) & (d) & (e) \\
\hline
$v$ $^{a}$ ($\rm km/s$)  & $[+400, +40]$  & $[+40, -320]$  & $[-320, -680]$  & $[-680, -1040]$  & $[-1040, -1400]$ \\
mean $^{b}$ & 0.10$\pm$0.05  & 0.11$\pm$0.05 & 0.56$\pm$0.04 & 0.43$\pm$0.05 & 0.27$\pm$0.05 \\
 EW$_1$ $^{c}$   &   &-0.31$\pm$0.21  &0.89$\pm$0.10   &$0.17\pm0.12$ \\
 EW$_2$ $^{d}$   &   &0.46$\pm$0.16   &0.26$\pm$0.16   &0.69$\pm$0.10\\
 |EW$_1$ $-$ EW$_2$|    &   &0.77$\pm$0.26   &0.63$\pm$0.18   & 0.51$\pm$0.15 \\
 $S\, ^{e}$   &   &3.0 $\sigma$   &3.5 $\sigma$   & 3.4 $\sigma$ \\
$ R_{\rm obs}=\frac{\rm EW_{weak}}{\rm EW_{strong}} \,$ $ ^{f}$  &   & &$\frac{\rm EW_2}{\rm EW_1} = 0.30\pm0.18$   & $\frac{\rm EW_1}{\rm EW_2} = 0.25\pm0.18$  
 \\
\hline
\hline
\end{tabular}
\tablefoot{
The \ion{Fe}{XXV} line profile was divided into five velocity intervals based on spectral binning. 
$^{a}$ Velocity interval.
$^{b}$ The mean EW derived from the time-averaged spectrum.
$^{c}$ The mean EW within gray shaded area, see Fig. \ref{fig_resolve_EW}.
$^{d}$ The mean EW within the blue-shaded area.
$^{e}$ The significance level of the variability, calculated with $|\rm EW_1 - EW_2|$.
$^{f}$ $R_{\rm obs}$ is the observed ratio.}
\end{table*}

\subsection{Time-resolved Resolve spectra and \ion{Fe}{XXV} analysis}
\label{parameter_set-up}

Each of the 19 time-resolved Resolve spectra was analysed over the same
five velocity intervals shown in Fig.~\ref{fig_5etaus}, within the
\ion{Fe}{XXV} energy range of $6.668$--$6.708~\mathrm{keV}$.
The local model parameters were configured according to the following assumptions.

We assume that the intensities of the narrow, broad and relativistic emission lines are constant in time. Allowing the intensity of the narrow and broad line components to vary does not affect the local continuum level near the \ion{Fe}{xxv} absorption line.
Since the amplitude of the relativistic line component is small and limited in energy band width, even potential variations in its intensity do not affect the derived continuum level more than a few percent and therefore can be ignored as far as the equivalent width determination of the absorption line is concerned.
We also assume that all absorption components remain constant during the observation, except for the \ion{Fe}{xxv} resonance absorption line.
We verified that even if some of the other absorption lines were variable, the effect on the derived continuum level at \ion{Fe}{xxv} is completely negligible.

Consequently, the covering factors, whose physical range is
  \(0\leq f_{\rm cov}\leq1\), are the only free parameter used to characterize the line absorption. 
A covering factor of zero means no absorption, whereas a covering factor of one corresponds to complete absorption within the velocity bin.
We formally allow for values smaller than zero or larger than one in order to obtain proper error estimates on the line depth. 
A negative value therefore does not imply emission or a negative physical covering factor; it means the measured flux lies statistically above the adopted local continuum.

\begin{figure*}[!tbp]
\centering
\hspace*{-0.00cm}\resizebox{0.85\hsize}{!}{\includegraphics[angle=0]{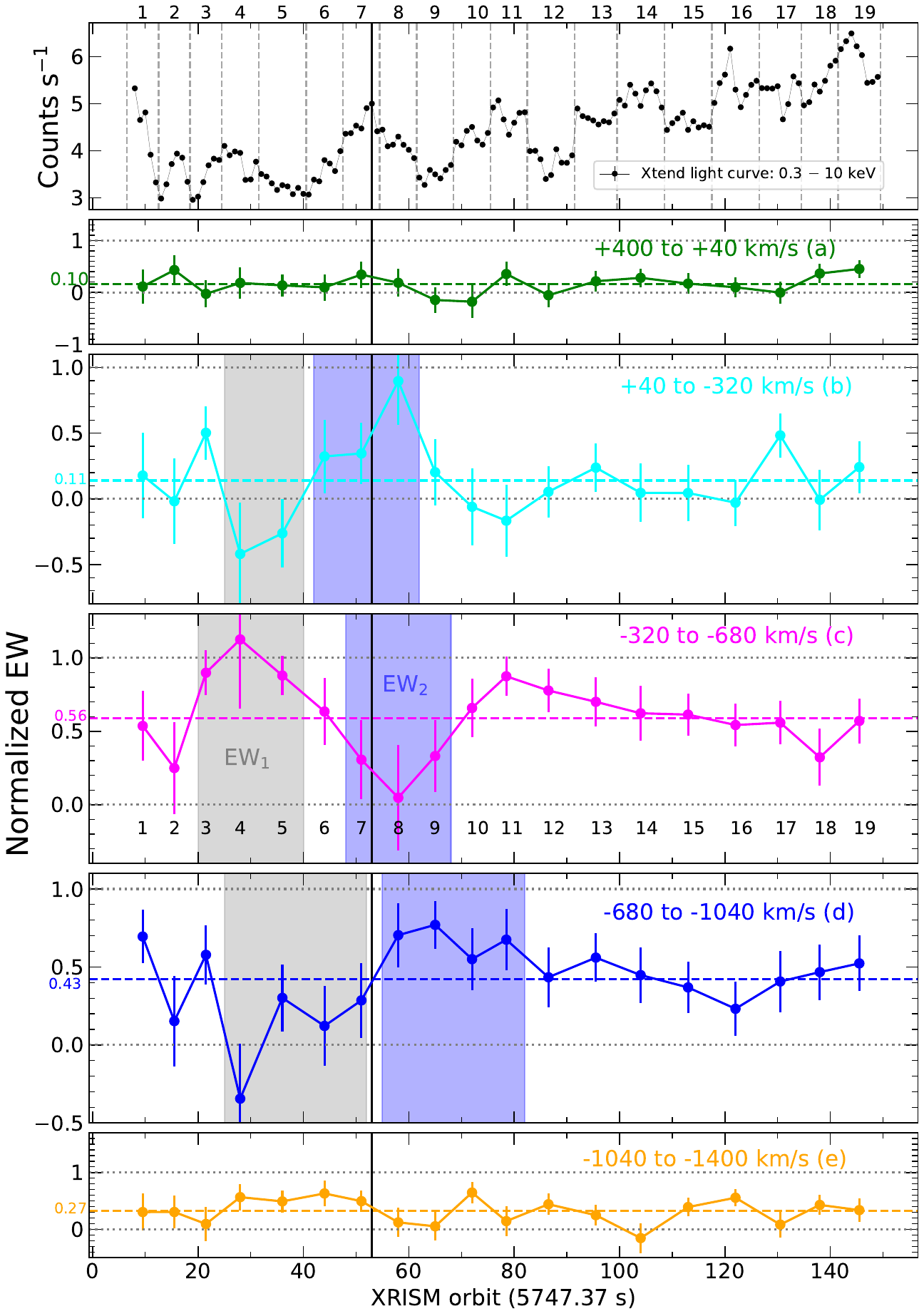}}\vspace{-0.0cm}
\caption{ 
Top: the Xtend light curve in the $0.3 - 10$ keV energy band for reference.
Panels (a) $-$ (e): the Resolve EW of the \ion{Fe}{XXV} absorption line as a function of time for five fixed velocity intervals in NGC 3783. 
The normalized EW, i.e. $w_{i}$ in each given time bin, is defined such that $w_{i}=0$ corresponds to no absorption, whereas $w_{i}=1$ corresponds to complete absorption within a bin. During fitting, $w_i$ is
allowed to extend beyond this physical interval to avoid boundary-biased
estimates.
The gray and blue shaded regions indicate the time intervals used to calculate the weighted means of normalized EW ($\rm EW_1$ and $\rm EW_2$) listed in the third and fourth rows of Table \ref{tab:19interval_stats}.
The colored dashed lines show the normalized EW measured from the time-averaged spectrum (see Sect. \ref{sect:local_fit_to_averaged}).
The black vertical line marks the peak time of flare C.
}
\label{fig_resolve_EW}
\end{figure*}

We then fit each time interval spectrum with the covering factor of the five \ion{Fe}{xxv} 
components and the photon index and normalisation of the main power law component as free parameters.
The five absorption components span the velocity range from $+400$ to $-1400$~km/s. 
The results of our fit for all line components are shown in Fig.~\ref{fig_resolve_EW}.

In this configuration, the fitted covering factor measures the
absorption depth within each velocity bin and is reported as the
normalized equivalent width (EW, hereafter), with its $1\sigma$
uncertainty.
Panels~(a)--(e) of Fig.~\ref{fig_resolve_EW} show the temporal evolution of
the \ion{Fe}{XXV} absorption in the five velocity intervals. The coloured
horizontal line in each panel marks the reference normalized EW measured
from the time-averaged spectrum. Values above and below this level represent
stronger and weaker absorption, respectively.
To quantify the variability, we calculated the weighted
mean normalized EW for selected phases of stronger and weaker absorption.
We summarize these results in Table \ref{tab:19interval_stats}.

\subsection{The observed \ion{Fe}{XXV} variability}
\label{obs_Fe25} 

Velocity interval~(c), spanning $-320$ to
$-680~\mathrm{km\,s^{-1}}$, covers the core of the
\ion{Fe}{XXV} absorption trough (Fig.~\ref{fig_5etaus}).
The \ion{Fe}{XXV} absorption strength varies from near-complete absorption (time interval 4) to negligible absorption (time interval 8) during 
the softest phase of the soft X-ray flare (Fig.~\ref{fig_resolve_EW}).
After reaching its minimum in interval 8, the EW increases until interval 11 and then shows a decreasing trend as the luminosity slowly increases.
The EW is high during orbits $20-40$ (gray panel in Fig. \ref{fig_resolve_EW}), with a value of $0.89 \pm 0.10$, whereas it decreases to $0.26 \pm 0.16$ during orbits $50-70$ (blue panel in Fig. \ref{fig_resolve_EW}). 
This corresponds to a difference of $0.63 \pm 0.18$, significant at the $3.5\, \sigma$ level.
This indicates that the \ion{Fe}{XXV} absorption line responds to the luminosity variation, especially around the soft flare event C.

In velocity interval (d), covering $-680$ to $-1040$ km/s, the blue wing of the absorption line, the EW exhibits significant variability around a mean value of EW = 0.43.
However, the amplitude is lower and it exhibits an opposite trend to that in interval (c).
In interval (b), for velocities $+40$ to $-320$ km/s, the EW evolves from no absorption (EW $\approx 0$) to nearly complete absorption within the velocity bin (EW $\approx1$) during flare C,
showing significant EW variability. However,
its apparent relationship with the continuum is uncertain because this
interval is strongly affected by the adjacent \ion{Fe}{XXV} emission
line. We therefore do not interpret its variability further.

In general, in Fig. \ref{fig_resolve_EW}, our analysis clearly shows that the \ion{Fe}{XXV} resonance absorption strength in the different velocity intervals exhibits different behaviours.
The EW in interval (d) reaches its maximum around the ninth time interval, whereas that in velocity interval (c) reaches its minimum at approximately the same epochs. 
The absorption in velocity intervals (c) and (d) shows opposite responses to the same variation in the light curve, suggesting contributions from physically distinct absorber components.
We note that interval (c), corresponding to the velocity range of $-320$ to $-680$ km/s and forming the deepest part of the \ion{Fe}{XXV} absorption trough, exhibits the most significant variability when the flux changes.

\section{Time-dependent photoionisation modelling: \textit{tpho}}
\label{sect3_model}

The observed variability of the \ion{Fe}{XXV} absorption line may be driven by changes in the illuminating X-ray flux. The response of the photoionized outflow depends on both the characteristic variability timescale of the incident light curve and the recombination timescale of the plasma (e.g., \citealt{Kaastra2012,Rogantini2022,Li2023}). 
The time-dependent photoionisation (\textit{tpho}\footnote{\url{https://spex-xray.github.io/spex-help/models/tpho.html}}) model implemented in SPEX v3.08.01 (\citealp{Kaastra2024}) tracks the ionisation evolution of the gas in response to X-ray luminosity variations (\citealt{Rogantini2022}).
To model the absorber response with \textit{tpho}, we adopt an SED, a driving light curve, and an initial photoionisation-equilibrium state
for each absorber.

\paragraph{SED.} NGC 3783 was observed in an unobscured state in 2024 comparable to that of 2001, with no evidence of a high-column-density obscurer along the line of sight (\citealp{Mehdipour2017}). 
For the 2024 XRISM observation, \citet{Mehdipour2025} adopted the 2001 SED to model the warm-absorber properties. For consistency, we use the same SED as \cite{Mehdipour2025} and refer to it as the ``normal SED''.
We neglect possible changes in the SED during flare C because the available data do not constrain its evolution reliably.

\begin{figure*}
    \sidecaption
    \includegraphics[width=12cm]{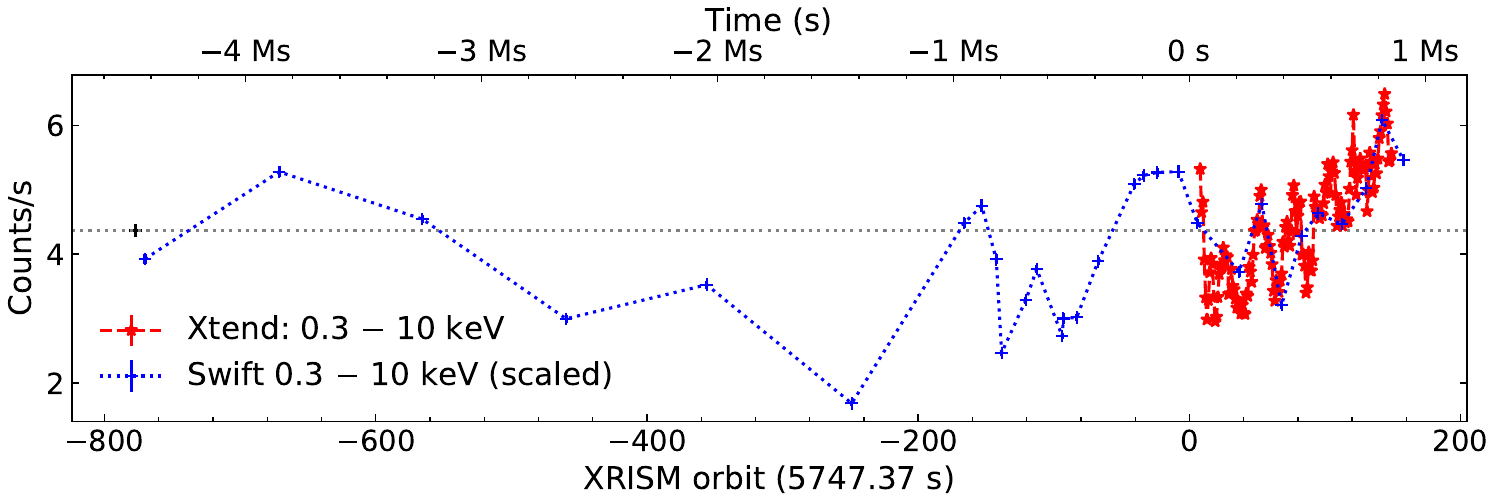}
    \caption{The combined Swift/XRT and XRISM/Xtend light curve in the energy band of $0.3-10$ keV.
The horizontal dashed line indicates the mean count rate of the combined light curve, adopted as the initial flux of the time-dependent calculation.}
    \label{fig_swfit_xtend_lc}
\end{figure*}

\paragraph{Light curve.}
We extracted XRISM/Xtend light curves in the $0.3$--$1.5$, $1.5$--$10$, and $0.3$--$10$ keV bands to match the corresponding Swift/XRT bands. 
In the same energy range, Swift/XRT provides longer-term monitoring, including seven epochs simultaneous with Xtend. 
We cross-calibrate and scale the Swift/XRT light curve to match Xtend by a scaling factor of  $3.2\pm0.5$, $3.0\pm0.8$ and $3.5\pm0.5$ for soft, hard and total band, respectively.
During flare C, we fitted the peak of the Xtend light curve with a linear baseline plus a Gaussian profile, yielding peak orbits of 52.6$\pm$0.2, 50.0$\pm$0.2, 51.2$\pm$0.1 in the soft (0.3$-$1.5 keV), hard (1.5$-$10 keV), and total bands (0.3$-$10 keV), respectively.

We plot the light curve in the energy band of $0.3 - 10 $ keV in Figure \ref{fig_swfit_xtend_lc}. We use it as input light curve for \textit{tpho} in our current calculation.
In the combined Xtend+Swift light curve, we exclude Swift data points that overlap in time with the Xtend coverage. 
We then take the mean count rate (the horizontal dashed line of Fig. \ref{fig_swfit_xtend_lc}) of the combined light curve as the first point of the \textit{tpho} driving light curve. 
We assume a photoionisation equilibrium state at the start of the calculations.

\paragraph{Initial photoionisation equilibrium state.}
We assume that the variability detected in the \ion{Fe}{XXV} absorption line (Sect.~\ref{sect_obs}) arises from changes in the physical absorber components identified in the time-averaged spectral analysis of \citet{Mehdipour2025}. 
\citet{Mehdipour2025} modelled the time-averaged Resolve spectrum with
six photoionised \textit{xabs} transmission components. For a specified
ionising SED, \textit{xabs} calculates the transmission of a slab of gas. 
Its principal fitted parameters include $N_{\mathrm{H}}$, $\log\xi$, the
outflow velocity, and the turbulent velocity (Table~\ref{tab:fe25_components_M}).
In that study, the \ion{Fe}{XXV} absorption features were attributed to three warm-absorber components: Comp.~A1, B, and C.
We take the parameters from the best-fitting time-averaged spectrum by \cite{Mehdipour2025} for this initial state.

\begin{table}
\centering
\caption{Adopted parameters of the \ion{Fe}{XXV} absorber components from the
time-averaged analysis of \citet{Mehdipour2025}, used as initial inputs
to the \textit{tpho} calculations.}
\label{tab:fe25_components_M}
\begin{tabular}{lcccc}
\hline\hline
Component &
$v_{\mathrm{out}}$ &
$\log \xi$ &
$N_{\mathrm{H}}$ &
$\sigma_{v}$ \\
&
($\mathrm{km\,s^{-1}}$) &
($\mathrm{nW\,m}$) &
($10^{25}\,\mathrm{m^{-2}}$) &
($\mathrm{km\,s^{-1}}$) \\
\hline
A1 & $-590$  & $3.19$ & $3.6$ & $40$  \\
B  & $-700$  & $2.96$ & $15$  & $420$ \\
C  & $-1170$ & $3.56$ & $9$   & $250$ \\
\hline
\end{tabular}
\end{table}

\subsection{Time-dependent \ion{Fe}{XXV} ion concentration calculations}
\label{tpho}
\begin{figure}[!tbp]
\centering
\hspace*{-0.00cm}\resizebox{1\hsize}{!}{\includegraphics[angle=0]{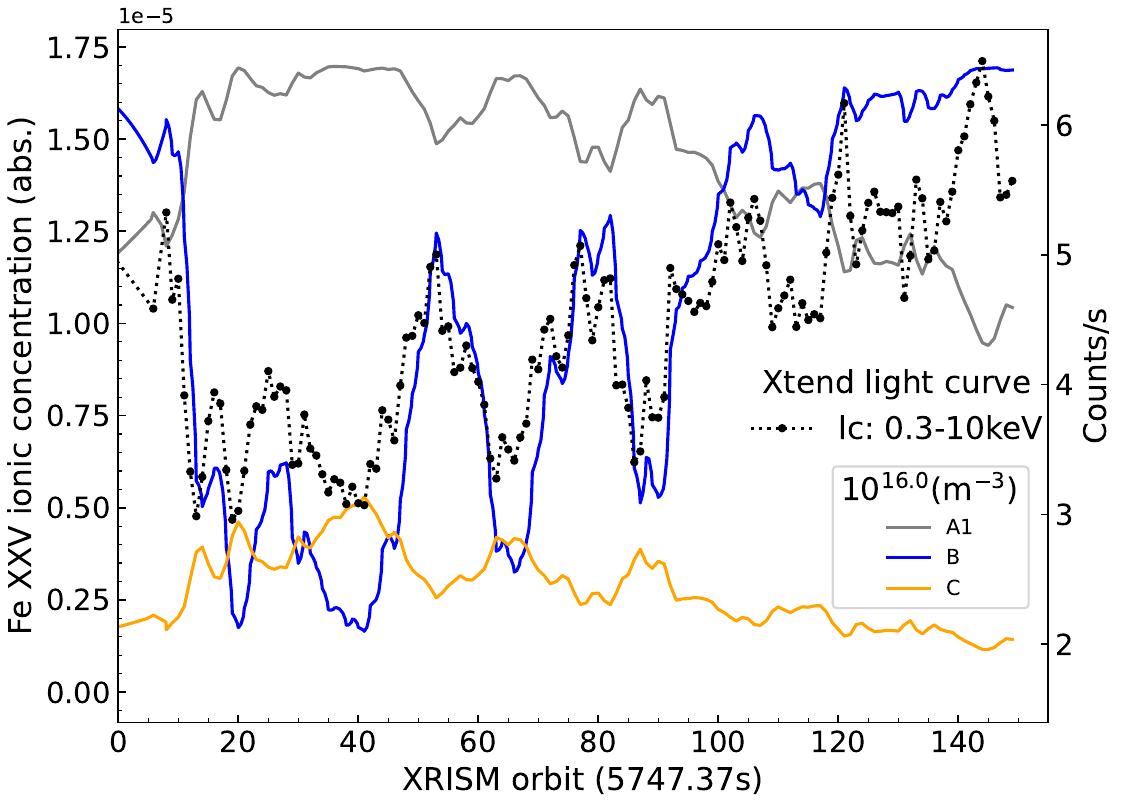}}\vspace{-0.0cm}
\caption{
Time-dependent \ion{Fe}{XXV} ion concentration predicted by \textit{tpho} for a gas density of $10^{16.0} \, \rm m^{-3}$. 
The gray, blue and orange curves represent Comp. A1, B and C, respectively.
The black dotted curve shows the input light curve from Fig. \ref{fig_swfit_xtend_lc}.
}
\label{fig_ionC}
\end{figure}

For the 
components A1, B, and C, we
performed separate \textit{tpho} calculations. Each calculation used
an energy grid spanning $0.01$--$10^8$ eV with a logarithmic step size of 0.1 dex.
The combined
$0.3$--$10~\mathrm{keV}$ light curve (Fig.~\ref{fig_swfit_xtend_lc}) drives the time-dependent
ionisation evolution over $5.2~\mathrm{Ms}$. For each component, the
corresponding $\log\xi$ and outflow velocity listed in
Table~\ref{tab:fe25_components_M} were used to define the initial equilibrium state.
Figure~\ref{fig_ionC} shows the predicted \ion{Fe}{XXV} ionic
concentrations of components A1, B, and C during the XRISM observation.
The Comp. A1 and Comp. C vary inversely with the light curve, whereas Comp. B responds positively to the light curve.
Comp. B shows a stronger temporal ionic response than Comp. A1 for the adopted initial ionisation states. Its larger column density and broader velocity profile also cause it to contribute strongly to the model EW in interval (c).

For each absorber component, the \textit{tpho} calculations are performed over a range of gas densities and yield the corresponding time-dependent \ion{Fe}{XXV} ionic concentration.
The resulting ionic concentration (hereafter, $C_{\rm ion} (t)$) was averaged for each of the 19 time intervals.
Using $N_{\rm ion} = N_{\rm H} \times C_{\rm ion}$, we convert the predicted $C_{\rm ion}$ into the ionic column $N_{\rm ion}$ for each component, where $N_{\rm H}$ is the hydrogen column density of the warm absorber component (see Table~\ref{tab:fe25_components_M}).
We insert these Fe ionic column densities in the \textit{slab} model of SPEX in order to compute the transmission of the absorber.
The \textit{slab}\footnote{\url{https://spex-xray.github.io/spex-help/models/slab.html}} model calculates the transmission through a thin
layer of gas from independently specified ionic column densities, a
velocity width $\sigma_v$, and a bulk velocity shift
\citep{Kaastra2002}.

\subsection{Comparison with the observed variability}
\begin{figure}[!tbp]
\centering
\hspace*{-0.00cm}\resizebox{1.0\hsize}{!}{\includegraphics[angle=0]{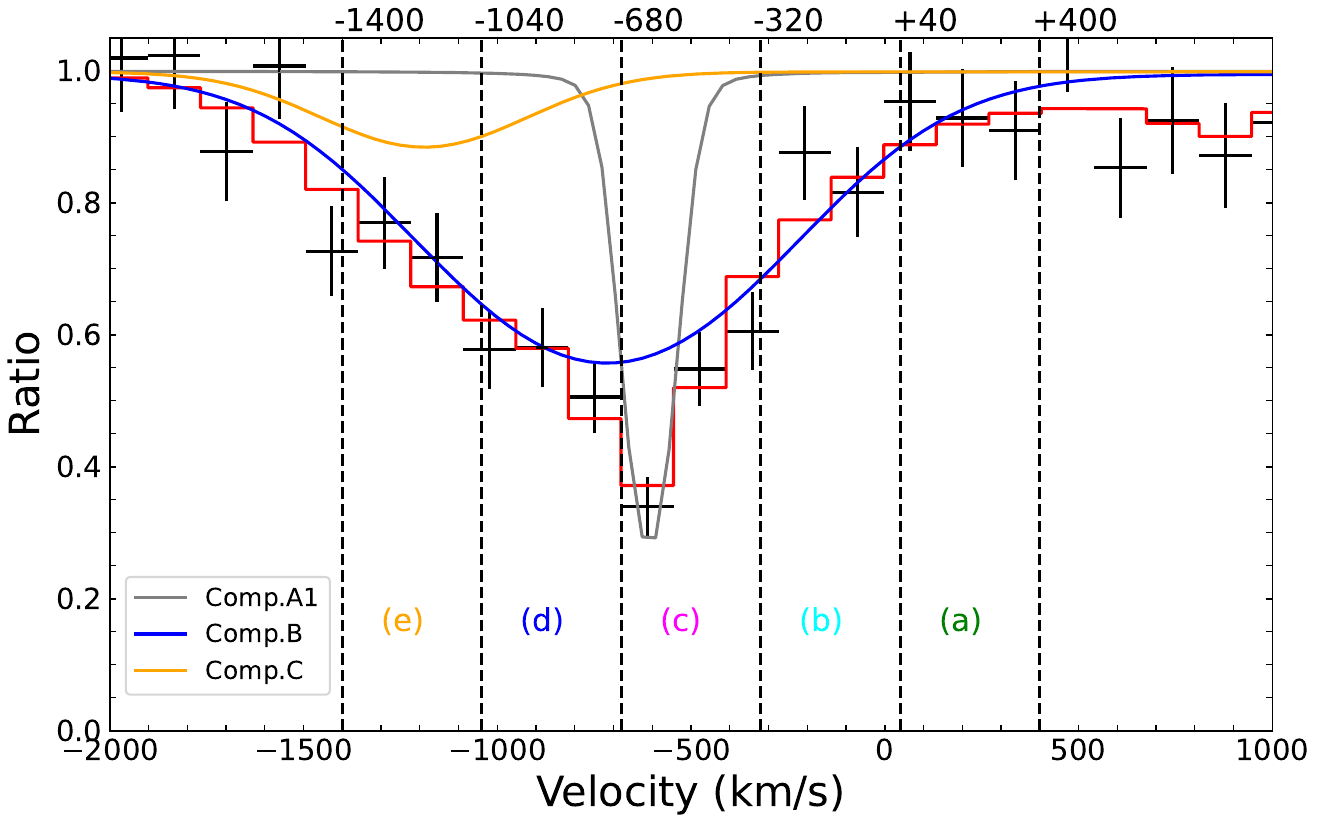}}\vspace{-0.0cm}
\caption{
The time-averaged spectral model (in red) of \cite{Mehdipour2025} identifies the contributions of
components A1, B, and C to the \ion{Fe}{XXV} absorption profile; their
best-fitting parameters are used to initialise the \textit{tpho}
calculations.
The vertical lines represent the same velocity intervals as in Fig.~\ref{fig_5etaus}. 
}
\label{fig_xabs}
\end{figure}

\begin{figure}[!tbp]
\centering
\hspace*{-0.00cm}\resizebox{1.0\hsize}{!}{\includegraphics[angle=0]{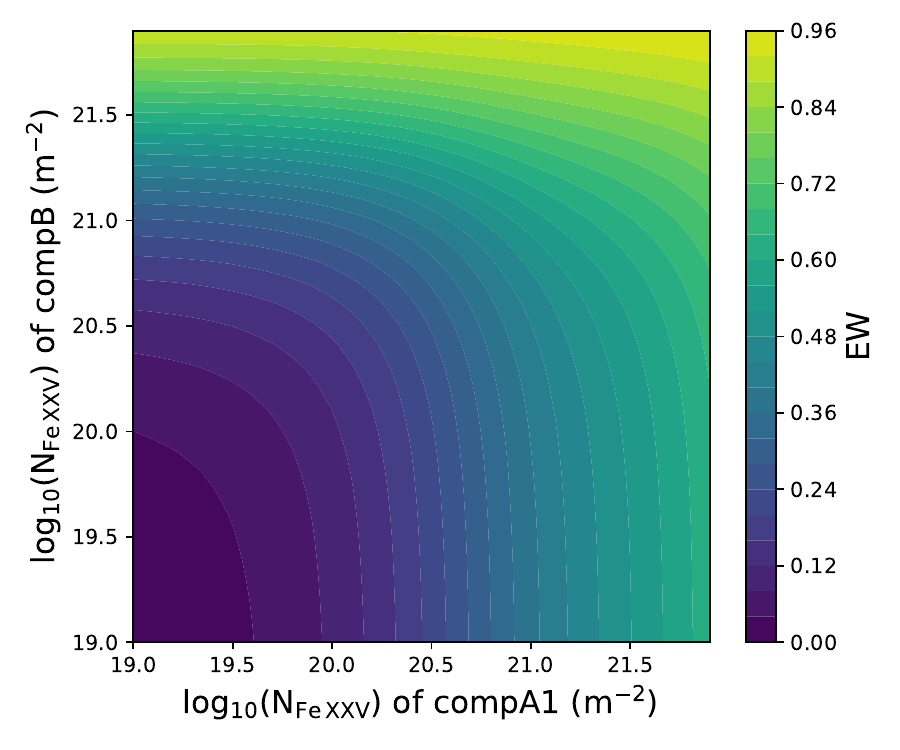}}\vspace{-0.0cm}
\caption{
The normalized model EW for velocity range (c) as a function
of the respective \ion{Fe}{XXV} column densities of Comp. A1
and B.
}
\label{fig_converter}
\end{figure}

We apply this procedure to velocity interval~(c), which exhibits the most
significant variability in Fig.~\ref{fig_resolve_EW}. As shown in
Fig.~\ref{fig_xabs}, absorption in this interval is dominated by components A1 and B. 
For a grid of values for the column densities of components A1 and B, we show the combined EW in Figure \ref{fig_converter}.
The combined EW increases with the ionic column density of either component.
At high column densities, however, the EW approaches unity and becomes less
sensitive to further increases in column density because the absorption line becomes saturated. 
Component B makes a substantial contribution to velocity interval~(c)
because of its relatively large column density and broad velocity
profile, which distributes its absorption across several adjacent
velocity intervals. 
However, a larger velocity width does not by itself
imply stronger saturation, since the line-centre optical depth scales
approximately as $N_{\rm ion}/\sigma_v$.
For each of the 19 time intervals, the pair
\(\left(N_{\ion{Fe}{XXV},\mathrm{A1}},
N_{\ion{Fe}{XXV},\mathrm{B}}\right)\) predicted by \textit{tpho} is
mapped onto this grid to obtain the corresponding model EW.

\begin{figure*}[!tbp]
    \centering
    \hspace*{-0.00cm}\resizebox{0.7\hsize}{!}{\includegraphics[angle=0]{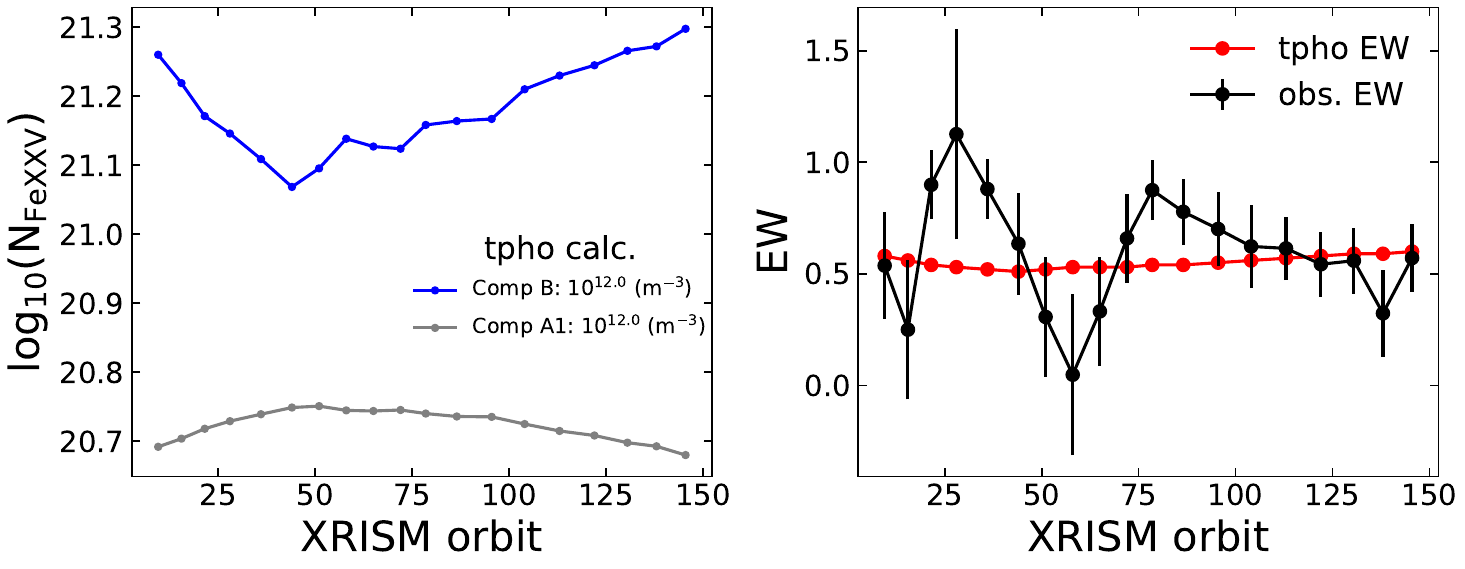}}\vspace{+0.5cm}
    \hspace*{-0.00cm}\resizebox{0.7\hsize}{!}{\includegraphics[angle=0]{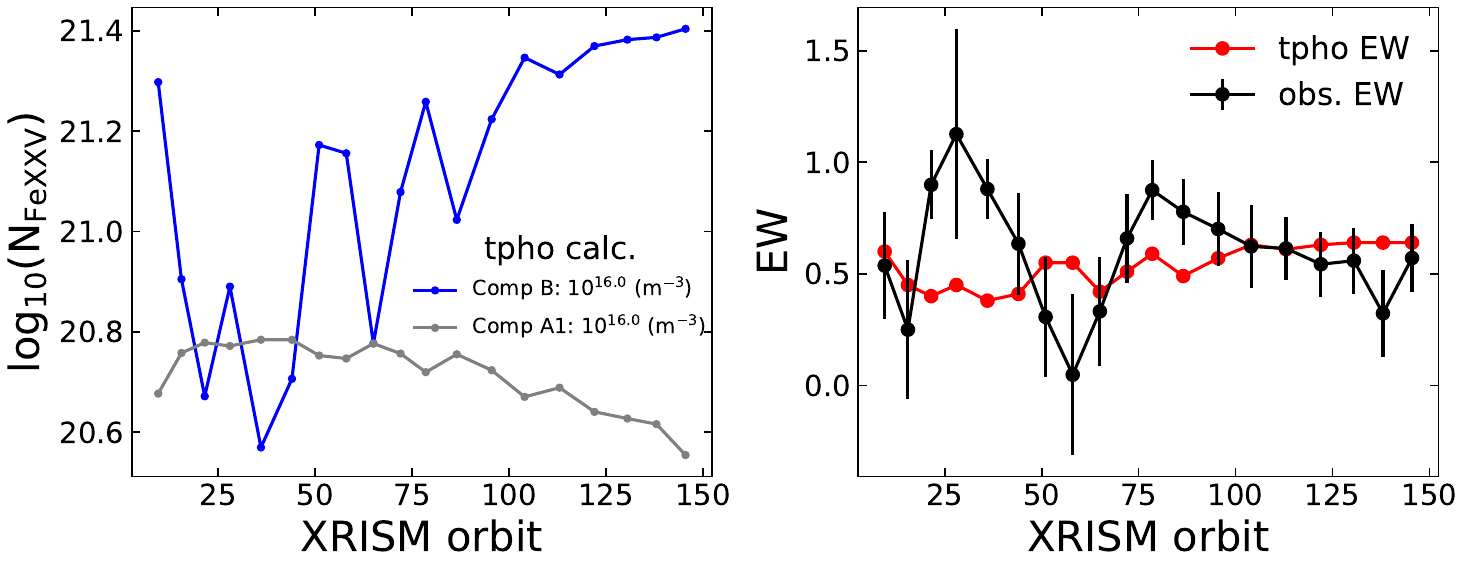}}\vspace{-0.0cm}
    \caption{
    Time evolution of the \ion{Fe}{XXV} absorption line for two density cases.
    Left panels: the calculated ionic column density in $\rm m^{-2}$ of \ion{Fe}{XXV} for Comp. B and A1. 
    Right panels: the combined EW of Comp.B and A1 for velocity range (c) as computed from the column densities in the left panels (red points and line), compared with the observed values (black).
    Top row: low density ($10^{12} \, \rm m^{-3}$).
    Bottom row: high density ($10^{16} \, \rm m^{-3}$).
    }
    \label{fig_model1_tpho}
\end{figure*}

Figure~\ref{fig_model1_tpho} shows the \ion{Fe}{XXV} ionic column densities
predicted by \textit{tpho} for components A1 and B over the 19 time
intervals (left panels) and compares the corresponding \textit{tpho} model EWs with the
observed EWs from Fig.~\ref{fig_resolve_EW} in velocity interval~(c) (right panels). In the lower-density
case (upper row), the long ionisation and recombination timescales produce
smooth variations in the ionic column densities. Consequently, the model EW
remains nearly constant and does not reproduce the observed variability.
In the higher-density case (lower row), component B responds more rapidly
to the ionising continuum and dominates the predicted EW variability,
whereas component A1 varies more slowly. Nevertheless, the high-density
model still does not reproduce the full amplitude of the observed
variability.
We explored different density combinations for these two components, but no satisfactory solution was found within the model, even in the high-density cases.

\section{Discussion}\label{sect_disc}

The time-resolved analysis of the \ion{Fe}{XXV} absorption line shows different behaviour for different velocity intervals.
As shown in Sect.~\ref{obs_Fe25}, the variability in velocity interval (c) is opposite to that observed in interval (d), suggesting that the absorption cannot be described by a single uniform component and that the narrow spectral-bin analysis traces changes in physically distinct absorber components. 
For this analysis, we treat intervals~(c) and (d) as tracing separate physical components.

\subsection{Limitations of our \textit{tpho} modelling}
\label{tpho_limit}
As noted in Section \ref{tpho}, our time-dependent photoionisation modelling with \textit{tpho} was unable to explain the observed spectral line variability. In this section, we discuss the possible reasons for this limitation.

First, the \textit{tpho} calculations assume that the SED retains the
shape measured during the unobscured 2001 epoch, while its normalisation
follows the observed light curve
(Fig.~\ref{fig_swfit_xtend_lc}). To compare the overall SED
normalisations in 2001 and 2024, we calculated the
$1$--$1000~\mathrm{Ryd}$
($13.6~\mathrm{eV}$--$13.6~\mathrm{keV}$) ionising luminosity using the
2001 normal-state SED of \citet{Mehdipour2017} and the best-fitting 2024
broad-band model of \citet{Zhao2026}. The two estimates agree within
their uncertainties, indicating comparable integrated ionising
luminosities.

Although HST and XMM-Newton constrain the UV and soft X-ray sides of the
ionising SED, respectively, no instrument in the campaign directly
covers most of the EUV continuum. Moreover, the temporal
coverage of HST and XMM-Newton is insufficient to reconstruct an
independent broad-band SED for each of the 19 time intervals
\citep{XRISMn3783cal2025}. We therefore adopt the well-constrained 2001
SED as a reference shape and vary its normalisation according to the observed light curve. Possible spectral-shape changes during flare C remain a systematic uncertainty in the \textit{tpho} calculations.
The energy-dependent evolution of the same flare, attributed to coronal magnetic reconnection \citep{Vardi2026}, suggests that scaling a fixed SED with a single light curve may not fully represent the ionising continuum.

Second, the \textit{tpho} predictions depend on the adopted initial
ionisation state and absorber decomposition. 
We initialise the calculations using the six-component solution of
\citet{Mehdipour2025}, whereas \citet{Zhao2026} identified eight
components from a joint Resolve and XMM-Newton/RGS analysis. 
The absorber
structure may therefore be more complex: component B may comprise
unresolved narrower components, while A1 is constrained mainly by one
spectral bin near $-600~\mathrm{km\,s^{-1}}$
(Fig.~\ref{fig_xabs}). Uncertainties in the initial ionisation
parameters, velocity widths, and component decomposition may thus
contribute to the
disagreement between \textit{tpho} prediction and the observation.
A broader profile
for A1 and a narrower profile for B could alter their relative
contributions to interval~(c),
however, exploring such a broad parameter space goes beyond the scope of this paper.

Finally, uncertainties in the underlying emission model
may affect the measured absorption variability.
The model includes a broad \ion{Fe}{XXV} emission line that affects the red wing of the absorption line \citep{Mehdipour2025}.  
The inclusion or exclusion of this emission line can also affect the fit to the time-averaged spectrum in the local energy band. The \ion{Fe}{XXV} emission may originate from the X-ray BLR \citep{Bianchi2002,Bianchi2005} and may vary significantly during the flare decay phase \citep{Gu2025}. 
The model also includes a separate relativistically broadened
emission component. 
A preliminary analysis by \citet{Li2026} suggests that
the equivalent width of this component varies by less than 30\%, except
during the strong soft flare. 
Because both emission components contribute
weakly in the local absorption band, their variability is
unlikely to dominate the measured absorption but is treated as a
systematic uncertainty.

\subsection{Density lower limit for the \ion{Fe}{XXV} absorbers}
\label{subsec:density}

The stronger variability of the \ion{Fe}{XXV} line is expected to occur for high densities.
For low densities, the line will remain constant in EW.
Therefore we focus here on high densities.
In this limit, at each time interval the ionisation state of the absorber is photoionisation equilibrium with the instantaneous time-dependent luminosity.

\subsubsection{Estimating the ionisation parameter from the EW ratio}
\label{sect:xil_EW}

In Table \ref{tab:19interval_stats}, we calculated the observed EW ratio for velocity intervals (c) and (d) to quantify the change in absorption strength during flare C. 
We did not calculate the ratio for interval (b), nor further constrain its $\log \xi$, because contamination from the \ion{Fe}{XXV} emission line introduces substantial uncertainty.
For a given absorber component in the optically thin regime, the equivalent
width is proportional to the ionic column density. If $N_{\mathrm{H}}$, the
covering fraction, elemental abundance, and velocity broadening remain
constant, then $N_{\mathrm{ion}}\propto C_{\mathrm{ion}}N_{\mathrm{H}}$.
Therefore,
\begin{equation}
\label{eq:EW_Nion}
    \frac{\mathrm{EW}_{\rm weak}}{\mathrm{EW}_{\rm strong}}
    =
    \frac{N_{\mathrm{ion, \, weak}}}{N_{\mathrm{ion, \, strong}}}
    =
    \frac{C_{\mathrm{ion, \, weak}}}{C_{\mathrm{ion, \, strong}}},
\end{equation}
where we compare two epochs with a weak and strong line.
This proportionality applies only in the optically thin regime.
Saturation will affect the stronger line more than the weaker, making this EW ratio bigger than in the optical thin limit.

\begin{figure}[ht!]
    \centering
    \includegraphics[
    width=1.0\linewidth,
    ]{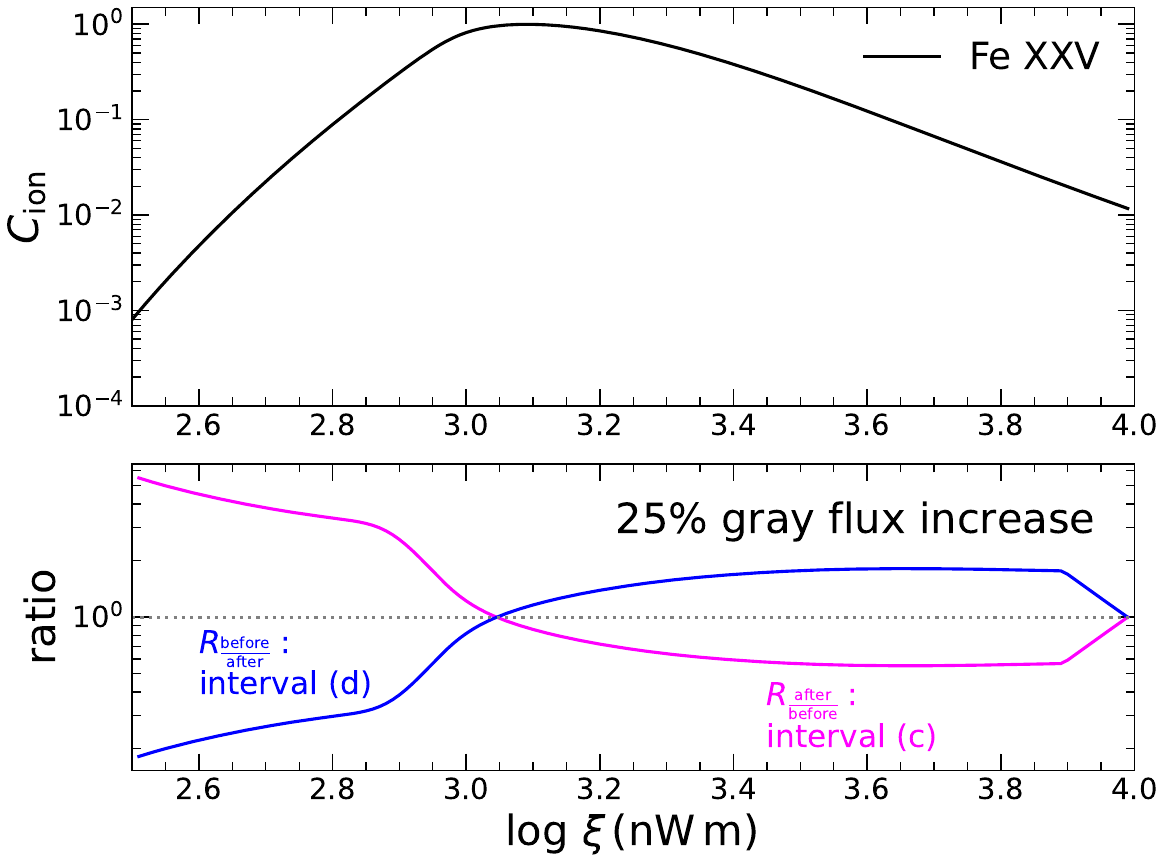}
    \caption{
    Top: Equilibrium \ion{Fe}{XXV} ionic concentration scaled to its peak value as a function of log$\xi$, calculated using the 2001 normal SED.
    Bottom: Ratio \(= C_{\mathrm{Fe\,XXV}}(\log\xi+0.10)/C_{\mathrm{Fe\,XXV}}(\log\xi)\) in magenta and the inverse in blue, both corresponding to a 25\% increase in ionising luminosity.
    }
    \label{fig_ionc}
\end{figure}

To constrain the ionisation parameter of the \ion{Fe}{XXV} absorbers,
we need to determine how the equilibrium Fe XXV ionic fraction varies with $\log\xi$.
The SPEX \textit{pion}
model\footnote{\url{https://spex-xray.github.io/spex-help/models/pion.html}}
calculates the equilibrium ionic fractions as a function of the ionisation
parameter \citep{Mehdipour2016}.
We run a grid of \textit{pion} models for different log$\xi$, using the 2001 unobscured SED. 
The resulting \ion{Fe}{XXV} ionic fraction distribution is presented in the top panel of Fig. \ref{fig_ionc}, peaking at log$\xi \approx 3.1$.
Since $\xi$ scales with luminosity as $\xi \propto L_{\rm ion} / (n_{\mathrm{H}}\,r^{2}) $, changes in $L_{\rm ion}$ will modify $\xi$ and therefore the \ion{Fe}{XXV} ionic fraction.
$L_{\rm ion}$ was enhanced by about $25\%$ during flare C corresponding to an increase of 0.10 in log$\xi$. 
The resulting $C_{\rm \ion{Fe}{XXV}}$ ratio of the before and after enhanced luminosity, defined as, 
\begin{equation}\label{eq:R_a_b}
R_{\frac{\rm after}{\rm before}} = \frac{C_{\ion{Fe}{XXV}}(\log\xi+0.10)}{C_{\ion{Fe}{XXV}}(\log\xi)},
\end{equation}
is plotted in Fig. \ref{fig_ionc}, together with its inverse i.e. $R_{\frac{\rm before}{\rm after}}$.

For comparing the observed EW ratios ($R_{\rm obs}$) with the model ratio, with the relationship shown in Eq.~\ref{eq:EW_Nion}, we defined the model EW ratio by $C_{\rm \ion{Fe}{XXV}}$ of weak to strong:
\begin{equation}\label{eq:model_EW}
R_{\rm model}(\log\xi)=
\frac{C_{\mathrm{ion, \, weak}}}{C_{\mathrm{ion, \, strong}}}. 
\end{equation}
Because the weak phase follows flare C in interval (c), its
observed ratio is compared with $R_{\rm \frac{after}{before}}$, shown
by the magenta curve in Fig.~\ref{fig_ionc}. 
In interval (d), the weak phase
precedes the strong phase, so its ratio is compared with
$R_{\rm \frac{before}{after}}$, shown by the blue curve.

For interval~(c), the measured ratio of $0.30\pm0.18$ is
compared with the declining magenta curve. The central and
lower values lie near or below the minimum model ratio and
therefore do not provide a unique constraint on $\log\xi$.
A significant decrease in the $C_{\rm \ion{Fe}{XXV}}$ begins at
$\log\xi\gtrsim3.2$ as the ionising flux increases. 
At $\log\xi\gtrsim3.5$, $C_{\rm \ion{Fe}{XXVI}}$ dominates over $C_{\rm \ion{Fe}{XXV}}$.
Requiring a detectable \ion{Fe}{XXV} fraction and significant
variability, we adopt $3.2\lesssim\log\xi\lesssim3.5$.

For interval~(d), the ratio of 
$0.25\pm0.18$ places the absorber on the rising blue ratio curve with log$\xi<3.05$. 
Its central ratio and upper limit derive log$\xi=2.65, 2.95$, respectively. 
Because the \ion{Fe}{XXV} fraction is negligible at the central value, we
adopt only the approximate upper limit $\log\xi\lesssim2.95$.

As indicated before, there may be some uncertainty in these limits on $\log\xi$ due to systematic effects. We have tested this for interval (c). Detailed spectral modelling of the line profile shows that due to the variability the \ion{Fe}{xxv} column densiy (see Eq.~\ref{eq:EW_Nion} decreases by a factor of $\sim$3 while the EW decreases only by a factor of $\sim 2$, but accounting for it does not change the derived limits on $\log\xi$ significantly. 
More importantly, we have adopted a change of about 25\% in luminosity during Flare C, corresponding to a change of 0.10 in $\log\xi$. However, when we inspect the light curve of Fig.~\ref{fig_xtend_interval}, we see that in the soft X-ray band (0.3–0.7~keV) the flux increases by a factor of more than two from the minimum at XRISM orbit 40 to the maximum at orbit 55. If we take that into account and repeat our analysis, the predicted EW ratio for interval (c) becomes completely in agreement with the observed ratio and constrains $\log\xi$ to a value larger than about 3.15. Most of the ionising luminosity in the spectrum of NGC~3783 is in the invisible EUV range, prohibiting a more detailed analysis. However, the above considerations do not alter the derived limits on $\log\xi$ in a serious way.

\subsubsection{Estimating the response timescales of the absorbers}

The transmission lags are measured relative to the X-ray continuum peaks because these peaks provide the driving changes in the ionizing radiation field. If the absorber responds through photoionization and recombination, variations in the transmission should follow the continuum changes with a characteristic response time set primarily by the recombination timescale \citep{Rogantini2022, Li2023}. 
The delay between the X-ray peak and the subsequent transmission response can be used to constrain the density of the absorbing plasma \citep{Li2025}.

We fitted the data in panels~(c) and (d) of
Fig.~\ref{fig_resolve_EW} locally with a linear baseline plus a
Gaussian. This determines $t_{\mathrm p}$, the time of the absorption
peak or trough in the blue-shaded region following flare C.
The $0.3$--$10~\mathrm{keV}$ light curve in Sect.~\ref{tpho} drives the \textit{tpho} calculations, whereas the $0.3$--$0.7$ and
$3$--$6~\mathrm{keV}$ light curves are used here to determine the
continuum-peak time and estimate the absorber-response lag. These
narrower bands largely avoid the Fe UTA absorption and Fe K emission
complexes, respectively, and therefore provide cleaner continuum
tracers. This lag estimate is independent of the \textit{tpho}
modelling. We define the lag as
\begin{equation}
    t_{\mathrm{lag}}=t_{\mathrm{p}}-t_{\mathrm{flare}},
\end{equation}
where $t_{\mathrm{flare}}$ is the continuum-peak time in the
corresponding energy band. Because neither band represents the full
ionising SED and the temporal sampling limits the accuracy of the
peak-time measurements, we estimate the lag relative to both the hard- and soft-band peaks (Fig.~\ref{fig_xtend_interval}) and adopt the larger upper bound as a conservative limit on the absorber response time.

The $3-6$ keV hard X-ray flux peaks at orbits 52, the $0.3-0.7$ keV soft X-ray flux peaks at orbit 55. 
The lag could be $5.5\pm3.5$ orbits (hard) or $2.5\pm3.5$ orbits (soft) for interval (c). Hence only an upper limit of $\sim 9$ orbits for interval (c) is realistic.
For interval (d), the peak of the response in the blue-shaded region of Fig.~\ref{fig_resolve_EW} is poorly constrained to the range of orbits $37 - 78$. 
The lag could be $5.2\pm20.6$ orbits (hard) or $2.2\pm20.6$ orbits (soft), which enables us to obtain an upper limit on the response time of $\sim 25.8$ orbits, or $\sim 150 \, \rm ks$.

\subsubsection{Density lower limits from the photoionisation-equilibrium approach}

We interpret the measured lag as an upper limit on the charge-state transition time, \(t_{\rm C}\), which describes the finite ionisation/recombination response of the plasma \citep{Li2023}. Therefore, \(t_{\rm C}\leq t_{\rm lag}\).
For velocity intervals (c) and (d), we obtain $t_{\rm C} \leq t_{\rm lag} \lesssim 50, 150 \rm \, ks$, respectively.
For a given SED, \(\log_{10}(t_{\rm C}n_{\rm e})\) is a function of
ionisation parameter and atomic number, 
which increases slightly with increasing log$\xi$ \citep{Li2023}.
The upper limit on $t_{\mathrm{C}}$ therefore gives a lower
limit on the electron density.

Using the normal-state 2001 SED and the same energy grid as
\cite{Li2023}, we obtain
a density lower limit for velocity interval~(c) of
$n_{\mathrm{e}}\gtrsim2.4\times10^{12}\ \mathrm{m^{-3}}$ for $\log\xi=3.2$, 
$n_{\mathrm{e}}\gtrsim4.24\times10^{12}\ \mathrm{m^{-3}}$ for $\log\xi=3.5$.
We therefore adopt the conservative lower limit
$n_{\mathrm{e}}\gtrsim2.4\times10^{12}\ \mathrm{m^{-3}}$
for interval~(c).
For interval (d), we assume log$\xi=2.95$ and obtain a lower density limit of  $n_{\rm e} \gtrsim 1.5\times10^{11} \, \rm m^{-3}$.
Because no lower limit on \(\log\xi\) is available for interval~(d), we cannot derive a more conservative density lower limit.
Electron-density limits were converted to hydrogen densities using the adopted $n_{\mathrm{e}}/n_{\mathrm{H}}$ ratio (\(1.2\) for highly ionised gas of near-solar composition).
These density constraints are summarized in
Table~\ref{tab:fexxv_flareC_panelc}.

\begin{figure*}[ht!]
    \centering
    \includegraphics[width=0.95\hsize]{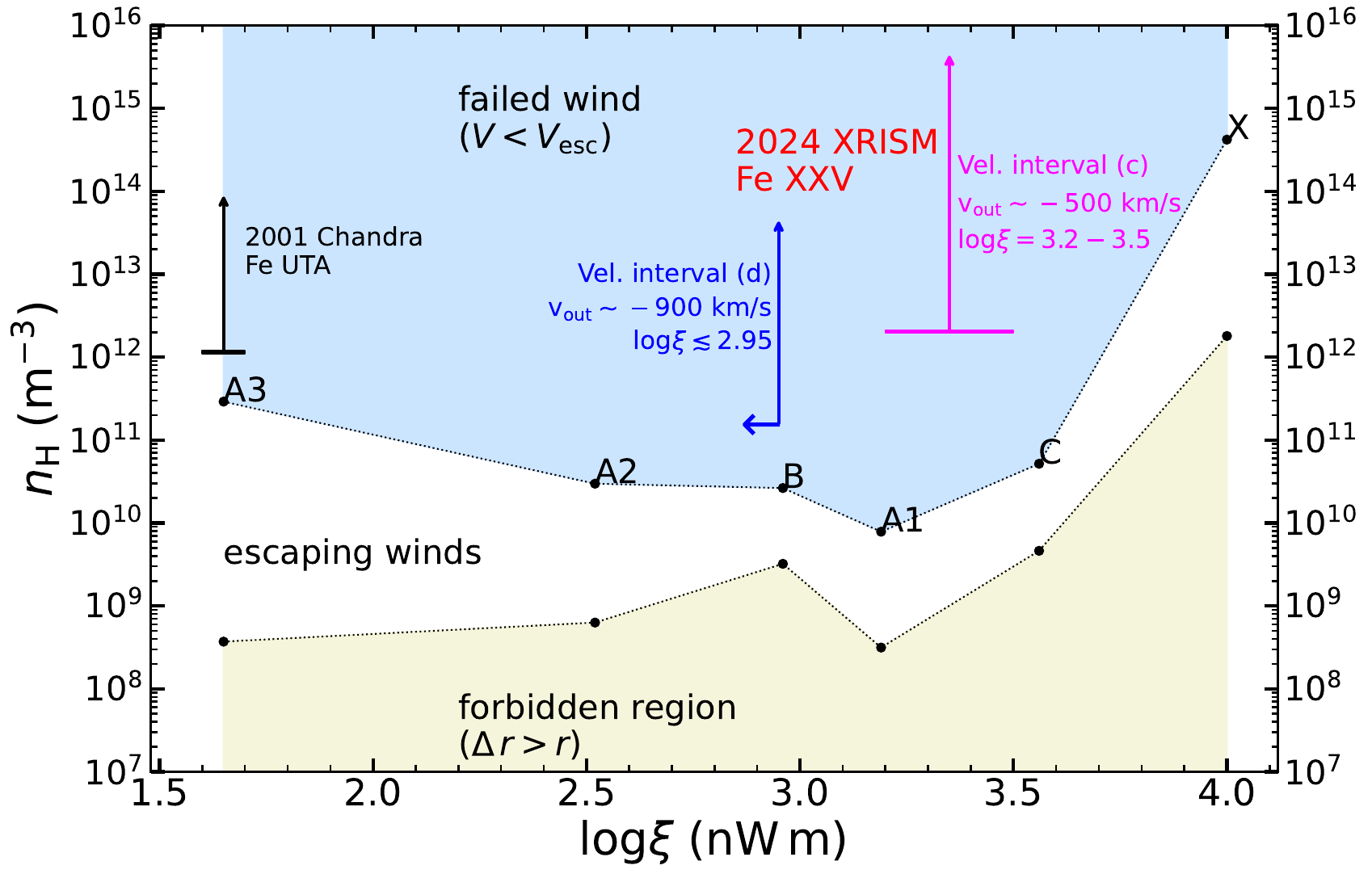}
    \caption{
    Hydrogen density as a function of ionisation parameter for the X-ray winds. The upper and lower density limits (black dots) for escaping winds are derived from the time-averaged spectrum \citep{Mehdipour2025}.
    The magenta and blue symbols show the results of the time-resolved analysis of \ion{Fe}{XXV} absorption variability in the present work. 
    The lower-ionisation failed wind detected with Chandra in 2001 is indicated by the black arrow \citep{Li2025}.  
    }
    \label{fig_ne}
\end{figure*}

Figure~\ref{fig_ne} compares the time-averaged XRISM density constraints with
the warm-absorber components identified with our current time-resolved spectra analysis and modelling. 
The upper boundary separates escaping and failed winds. Requiring
$v_{\mathrm{out}}\geq v_{\mathrm{esc}}$ gives
\begin{equation}
    n_{\mathrm{H}}\leq n_{\mathrm{H,esc}}
    =\frac{L_{\mathrm{ion}}v_{\mathrm{out}}^4}
    {4\xi(GM_{\mathrm{BH}})^2}.
\end{equation}
Gas above this boundary has $v_{\mathrm{out}}<v_{\mathrm{esc}}$ and is
therefore a failed wind. The lower boundary follows from
$\Delta r\leq r$ and $N_{\mathrm{H}}\simeq n_{\mathrm{H}}\Delta r$ \citep{Krolik2001, Blustin2005}:
\begin{equation}
    n_{\mathrm{H}}\geq n_{\mathrm{H,geom}}
    =\frac{\xi N_{\mathrm{H}}^2}{L_{\mathrm{ion}}}.
\end{equation}
Densities below this boundary would imply $\Delta r>r$ and are therefore
geometrically forbidden. The dotted lines connect the component-specific
limits and serve only as guides to the eye.

We find that the highly ionized absorber interval (c) is located within the failed-wind region, while
the interval (d) is also a failed wind candidate, although an escaping-wind solution cannot be excluded.

\subsection{Distances and energetics of the highly ionised \ion{Fe}{XXV} absorbers}
\label{}

\begin{table}[t]
\caption{
Parameters of velocity intervals (c) and (d) derived from the absorption line variability. 
}
\label{tab:fexxv_flareC_panelc}
\centering
\begin{tabular}{l c c}
\hline\hline
\multicolumn{3}{c}{Observed quantities} \\
\hline
Velocity interval &  (c)  &  (d)  \\
\hline
$v \, ^{a}$ ($\mathrm{km~s^{-1}}$) &  $[-320, -680]$  & $[-680, -1040]$  \\
$t_{\rm p}$ $^{b}$ (orbits)  &  57.5$\pm$3.5    & 57.2$\pm$20.6   \\
$t_{\rm lag} \, ^{c} \, (\rm ks)$   & < $50$    & $< 150$   \\
\hline
\multicolumn{3}{c}{Predicted quantities} \\
\hline
$\log\xi$ $^{d}$ &  $3.2 - 3.5$   & $\leq2.95$  \\
$n_{\rm H} \, ^{e} \, (\mathrm{10^{11} \, m^{-3}})$  & $> 20$  & $> 1.2$  \\
$r \, ^{f} \, (\mathrm{pc})$  & $< 0.05$    & $< 0.25$   \\
$v_{\rm esc}$ $^{g}$ ($\rm 10^3 \, km/s$) & $> 2.4$  &$> 1.0 $  \\
\hline
\end{tabular}
\tablefoot{
$^{a}$ The kinematic velocity range. 
$^{b}$ peak or trough time (orbits).
$^{c}$ lag timescale.
$^{d}$ Estimated log$\xi$ from Fig.~\ref{fig_ionc} (see Sect.~\ref{sect:xil_EW}). 
$^{e}$ Lower limit hydrogen density.
$^{f}$ Upper limit distance.
$^{g}$ Lower limit escape velocity. 
}
\end{table}

With the density and ionisation parameter derived for the absorbers in intervals (c) and (d), we can further estimate their distances from the SMBH. 
We summarize the derived properties of these two highly ionized \ion{Fe}{XXV} components in Table~\ref{tab:fexxv_flareC_panelc}.

For interval (c), we substitute $L_{\rm ion}$, log$\xi = 3.2$, and $n_{\rm H} = 2.0\times10^{12}\,\rm m^{-3}$ into Eq.~\ref{equ:xi defination}. 
Here, the 1--1000~Ryd ionizing luminosity is taken to be  $L_{\rm ion} = 6.4 \times 10^{36}\,\rm W$ from the broad-band continuum modelling of \cite{Mehdipour2017}. This yields an upper limit of \(r<0.05\) pc for interval (c).
To examine whether this absorber is gravitationally bound, we take $M_{\rm BH} = 3.0\times10^7\,M_{\odot}$ \citep{Bentz202101a} and $r = 0.05 \,\rm pc$ to calculate the local escape velocity. 
This gives $v_{\rm esc} > 2.37\times10^{3}\,\rm km\,s^{-1}$, which is significantly larger than the observed velocity range of $-320$ to $-680\,\rm km\,s^{-1}$. This result suggests that the absorber may therefore be gravitationally bound to the SMBH.

Given log$\xi=2.95$ and $n_{\rm H} = 1.22 \times 10^{11} \rm \, m^{-3}$ for interval (d), the derived distance upper limit is \(r< 0.25\) pc, while the resulting lower limit on local escape velocity is $1.02\times10^3 \, \rm km/s$, comparable with the observed outflow speed and favouring a plausible failed-wind interpretation.
However, because no lower bound on \(\log\xi\) is available, an escaping solution cannot be excluded for interval (d) over the full allowed range.

We also estimate the kinetic luminosity of the outflow using
\begin{equation}\label{eq:lkin}
    L_{\rm kin} = \frac{1}{2}\dot{M}_{\rm out} v_{\rm out}^2
    = \frac{1}{2}\mu m_{p} n_{\rm H} r^2 \Omega C_{\rm v} v_{\rm out}^3,
\end{equation}
where $\dot{M}_{\rm out}$ is the mass outflow rate, $\mu$ is the mean atomic weight per proton ($\approx 1.43$), $m_{p}$ is the proton mass, $\Omega$ is the solid angle, $C_{\rm v}$ is the volume filling factor, and $r$ is the radial distance from the source. 
Adopting a fiducial solid angle of $\Omega \sim 2\pi$ \cite{}, and $C_{\rm v} \sim 1$ under the assumption of a homogeneous and volumetrically thick outflow, taking into account the averaged outflowing velocity of $|v_{\rm out}| = 500$ and $900\,\rm km\,s^{-1}$ for intervals (c) and (d), respectively,
we derive the kinetic luminosities of $L_{\rm kin} = 3.8 \times 10^{33}$ and $3.9 \times 10^{34}$ $\rm W$.

Using the bolometric luminosity $L_{\rm bol} = 2.1 \times 10^{37}\,\rm W$ from the broad-band continuum modelling of \cite{Mehdipour2017}, corresponding to an Eddington ratio of $L_{\rm bol}/L_{\rm Edd} = 0.06$, we obtain $L_{\rm kin}/L_{\rm bol} = 0.0002$ and $0.0019$ for intervals (c) and (d), respectively. 
Under the adopted values of $\Omega$, $C_v$, and $\xi$, these
fiducial kinetic powers are too small to provide significant
feedback to the host galaxy.
The higher-velocity component has a less restrictive distance upper limit of \(r<0.25\) pc; its actual distance relative to interval (c) ($r<0.05 \rm pc$) cannot be determined from the present limits.
Using XMM-Newton/PN spectra, \cite{Reeves2004} observed significant \ion{Fe}{XXV} line variability within $10^{5}\,\rm s$, with the absorption strength becoming stronger at a relatively high flux level. 
The \(<0.1\)-pc constraint of \cite{Reeves2004} is consistent with the distance limits derived here.

\subsection{Possible origins of the observed \ion{Fe}{XXV}
variability in NGC~3783}

The opposite responses following flare C---weaker absorption in
interval (c) and stronger absorption in interval (d)---qualitatively
support two absorbers lying on opposite sides of the equilibrium
\ion{Fe}{XXV} ion-fraction peak (Fig.~\ref{fig_ionc}). Although
\textit{tpho} does not reproduce the full variability amplitude in
interval (c), this may reflect uncertainty in the initial equilibrium
ionisation state rather than rule out a photoionisation response
(Sect.~\ref{tpho_limit}).

An alternative explanation is transverse motion of inhomogeneous
absorbing gas across the line of sight \citep{Kraemer2001a, Crenshaw2003}. 
Considering the possible transverse motion of the \ion{Fe}{XXV} outflow across an X-ray emitting region of $\leq 20 \,\rm R_{g}$ \citep{Reis_Miller2013}, the crossing time is defined as \citep{Li2023}
\begin{equation}\label{equa:t_cross}
    t_{\rm cross} = \frac{20}{c^2} \, (G \times M_{\rm BH})^{1/2} \, \left(\frac{L_{\circ}}{\xi_{\circ}}\right)^{1/4} \, n_{\rm H}^{-1/4} \, ,
\end{equation}
where $L_{\circ} = 6.36 \times 10^{36} \,\rm W$ \citep{Mehdipour2017} and $\xi_{\circ}$ is the
adopted ionisation parameter of the corresponding absorber. 
At the density lower limits, we obtain
$t_{\rm cross}\lesssim\,  5.3 \, \times10^{5}$ and
$\lesssim\, 1.23 \, \times10^{6}~\mathrm{s}$ for intervals~(c) and (d),
respectively. 
These timescales are approximately 11 and
8 times longer than the corresponding response-time upper limits of
50 and 150~ks.

Changes in the geometry or surface-brightness distribution of the
X-ray corona could also alter the effective sight lines through an
inhomogeneous absorber, producing apparent changes in absorption
depth \citep{WilkinsGallo2015, Risaliti2009, Jiang2022}.
The newly detected relativistic Fe fluorescence line may provide
additional constraints on the disc--corona geometry
\citep{Li2026}.
Local density fluctuations provide another possible
explanation \citep{Waters2017,Mizumoto2019}. 
However, neither scenario naturally explains the opposite line
responses associated with flare C.
Longer XRISM observations with
multiple flares and simultaneous broadband coverage are needed to
distinguish these scenarios from photoionisation-driven variability.

\section{Summary}
\label{sect_conclusion}

We present a time-resolved analysis of the \ion{Fe}{XXV} absorption line observed during the 10-day XRISM/Resolve campaign of NGC~3783. 
Three velocity intervals exhibit significant equivalent-width (EW) variability. 
Intervals (c) and (d) respond differently between the low- and high-flux states, indicating temporal changes in the physical absorber components. 
However, direct time-dependent photoionisation modelling of the lower-velocity component with \textit{tpho} cannot reproduce the strongest EW variation. 
This is likely because the initial equilibrium solution derived from the time-averaged spectrum is not unique. 
As a result, we are unable to obtain a precise density constraint for this component.

Using photoionisation calculations (\textit{pion}) based on the 2001 unobscured SED and assuming the high-density limit, we estimate the ionisation-parameter ranges of these two velocity components. The lower-velocity component, at $-320$ to $-680~\rm km\,s^{-1}$ with $\log \xi \sim 3.2-3.5$, shows an inverse response to the continuum with a response-time upper limit of $\sim 50$ ks, implying a density of $n_{\rm e}> 2.38 \times 10^{12}~\rm m^{-3}$ and a distance of $< 0.05 $ pc.
The higher-velocity component, at $-680$ to $-1040~\rm km\,s^{-1}$ with $\log \xi \lesssim 2.95$, shows a
positive response with a lag upper limit of approximately
$150~\mathrm{ks}$.
This implies a density lower limit of $n_{\rm e} > 1.46 \times 10^{11}~\rm m^{-3}$, and a distance upper limit of \(r<0.25\) pc.
The absorber associated with interval (c) is a failed-wind candidate because its outflow velocity is below the local escape velocity. 
The higher-velocity interval (d) may lie near the failed wind threshold. 
Together with the failed Fe UTA wind identified in 2001, these results suggest that NGC~3783 hosts a multiphase failed-wind structure in its unobscured state.

\begin{acknowledgements}
C.L. acknowledges support from the Chinese Scholarship Council (CSC) and Leiden University/Leiden Observatory, as well as SRON.
SRON is supported financially by NWO, the Netherlands Organization for Scientific Research.
%
M. Mehdipour acknowledges support from the NASA XRISM Guest Scientist (XGS) grant (80NSSC23K0995).
%
S. Ogawa acknowledges JSPS KAKENHI grant No. 24K17104.
\end{acknowledgements}

\bibliographystyle{aa}
\bibliography{references}

\clearpage

\end{document}